\documentclass{article}
\usepackage[utf8]{inputenc}
\usepackage{graphicx}
\usepackage{authblk}
\usepackage[a4paper]{geometry}
\begin{document}

\title{Ion source and LEBT of KAHVELab proton beamline}

\author[1,3]{A. Ad{\i}g\"{u}zel}
\affil[1]{\.{I}stanbul University, Department of Physics, \.{I}stanbul}
\author[2]{S. A\c{c}{\i}ks\"{o}z}
\affil[2]{Bo\u{g}azi\c{c}i University, Department of Physics, \.{I}stanbul}
\affil[3]{Bo\u{g}azi\c{c}i University, Feza G\"{u}rsey Center for Physics and Mathematics, \.{I}stanbul}
\affil[4]{Y{\i}ld{\i}z Technical University, $^a$Electronics and Communication Engineering, $^b$Department of Physics, $^c$ Department of Control and Automation, \.{I}stanbul}
\author[4,a]{A. \c{C}a\u{g}lar}
\author[5]{H. \c{C}etinkaya}
\affil[5]{K\"{u}tahya Dumlup{\i}nar University, Department of Physics, K\"{u}tahya}
\author[1]{\c{S}. Esen}
\author[4,b]{D. Halis}
\author[1]{A. Hamparsuno\u{g}lu}
\author[4,c]{T.B. \.{I}lhan}
\author[6]{A. K{\i}l{\i}\c{c}gedik}
\affil[6]{Marmara University, Department of Physics, \.{I}stanbul}
\author[1]{O. Ko\c{c}er}
\author[7]{S. O\u{g}ur}
\affil[7]{Irene Joliot-Curie Lab., University of Paris-Saclay, Paris}
\author[2]{S. \"{O}z}
\author[8]{A. \"{O}zbey}
\affil[8]{\.{I}stanbul University Cerrahpa\c{s}a, Department of Mechanical Engineering, \.{I}stanbul}
\author[2,3]{V.E. \"{O}zcan}
\author[9]{N.G. \"{U}nel}
\affil[9]{University of California Irvine, Physics Department, Irvine}
\date{ \today }
\maketitle

\begin{abstract}
The KAHVE Laboratory, at Bo\u{g}azi\c{c}i University, Istanbul, Turkey is home to an educational proton linac project. The proton beam will originate from a 20\,keV H+ source and will be delivered to a two module Radio Frequency Quadrupole (RFQ) operating at 800\,MHz via a low energy beam transport (LEBT) line. 
Currently, the design phase being over, commissioning and stability tests are ongoing for the proton beamline which is already produced and installed except the RFQ which is being manufactured. This work summarizes the design, production and test phases of the ion source and LEBT line components. 
\end{abstract}

\section{Introduction}

Kandilli Detector, Accelerator and instrumentation laboratory (KAHVELab) started the construction of a proton beamline funded by TUBITAK\footnote{Scientific and Technological Research Council of Turkey}, Bogazici and Istanbul Universities.  The project aims to build a Proton Testbeam At Kandilli (named as the PTAK project) campus in Istanbul, Turkey. The PTAK project is being conducted based on previous experiences \cite{SPPpaper} and with the main goal of educating the next generation of accelerator physicists and engineers on the job and accumulating operational know-how. To attain this goal, most of the components are locally designed and manufactured in tandem with local companies. 
A secondary purpose of this project is to be a particle accelerator technologies test setup with the ultimate goal of providing beamtime to other projects such as Proton Induced Xray Emission (PIXE) measurements. 

The PTAK beamline design consists of an ion source, a low energy beam transport (LEBT) section and a radio frequency quadrupole (RFQ). Such a simple setup is the first stage of almost all modern ion accelerators such as the Large Hadron Collider at CERN. Low energy ion machines have fast-moving applications in the field of science and technology such as accelerator physics, atomic physics, plasma physics and chemistry, nuclear physics, mass spectroscopy, isotope separation, controlled thermonuclear fusion, radiation chemistry, ion inoculation, microanalysis and microfabrication. Benefiting from accelerator physics and its technologies, it is possible to conduct studies which will bring innovation to science and technology in many fields such as material science, medical physics and nuclear energy. Due to its applicability to these areas, proton machines are more frequently seen than heavier ions. 

\begin{figure}[!htbp]
\centering
\includegraphics[scale=0.30]{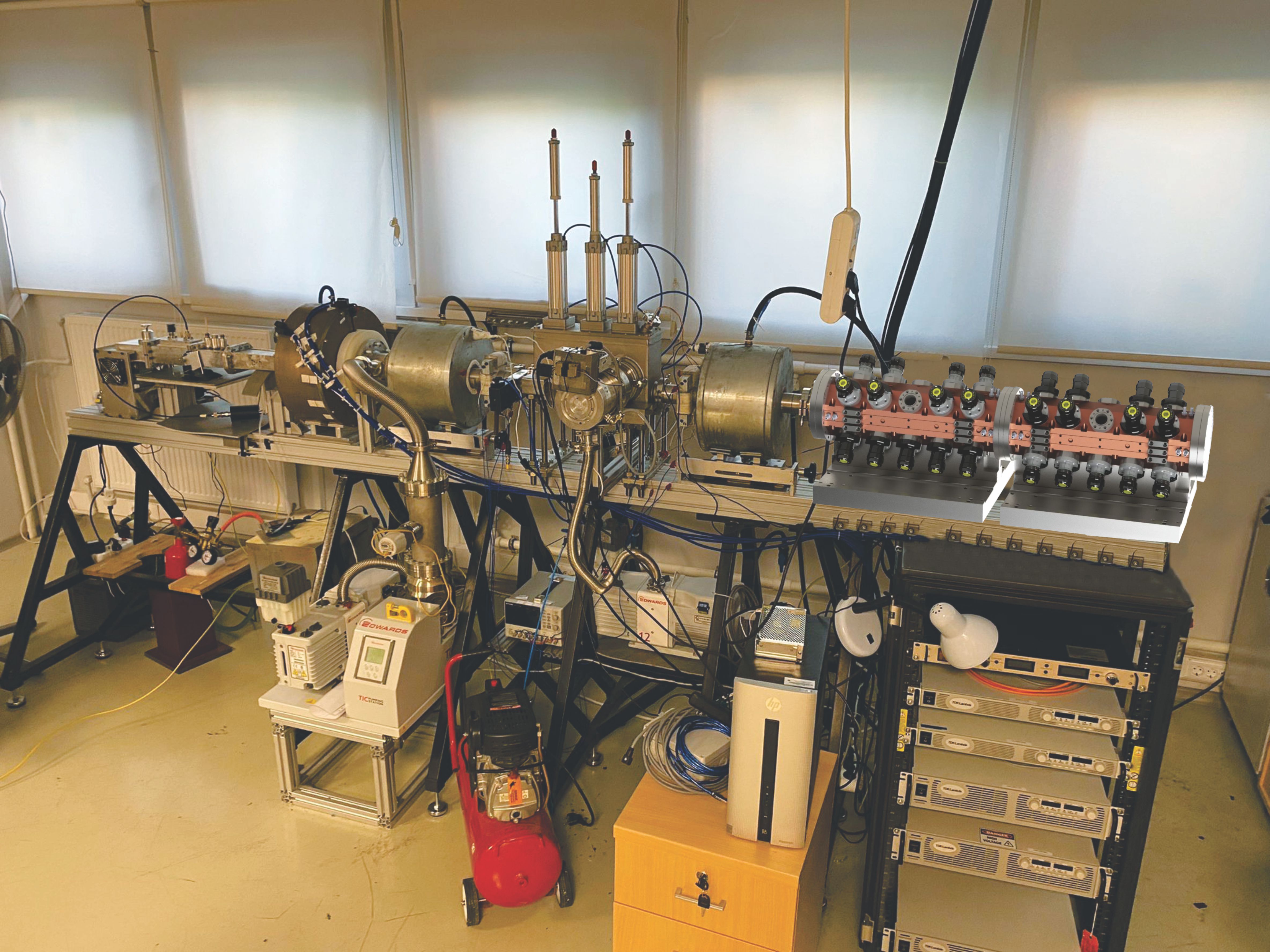}
\caption{The Proton Beamline at KAHVELab. The RFQ is included virtually in this photo.}
\label{fig:beamline}
\end{figure}

Currently,  the PTAK design is completed, the ion source and the LEBT line is constructed, installed and are being commissioned. The RFQ's test module is also constructed and it's electromagnetic and vacuum tests are being conducted \cite{RFQModule0}. The PTAK beamline is expected to be fully commissioned by the end of 2023. A view of the KAHVELab proton beamline installation can be seen in Fig. \ref{fig:beamline}. From left to right, the main components are: the RF power source, the ionization chamber, the first solenoid, the measurement station and the second solenoid. The missing component, Radio Frequency Quadrupole, is added virtually. 

The PTAK design is optimized for a low current (maximum instantaneous current of 1mA) and low duty factor pulsed beam (currently 20\%, to be lowered furthermore) which would be accelerated from 20~keV to about 2~MeV energy in about 1~m distance. This feat will be achieved by building a four-vane type RFQ operating at 800~MHz, which will be the highest frequency machine in operation. The beamline parameters are summarized in Table \ref{tab:design}.

\begin{table}
\caption{PTAK beamline global parameters.}
\begin{center}
\begin{tabular}{|c|c|c|}
  \hline
  Parameters & Value & Unit \\ \hline 
  p extraction & 20 & kV  \\  \hline
  I$_{peak}$ & 1 & mA       \\  \hline
  d.f. & $<$ 20\% & -        \\  \hline 
  LEBT length & 165 & cm \\ \hline
  RFQ length & 98 & cm    \\  \hline
  E$_{out}$ & 2 & MeV \\ \hline
  RFQ frequency & 800 & MHz    \\
  \hline 
\end{tabular}
\label{tab:design}
\end{center}
\end{table}

The rest of this paper discusses on the ion source and LEBT design, construction and commissioning and omits the details on how the measurements were taken. The details of the measurement station within the LEBT can be found elsewhere \cite{MBOXPaper}. The details of the RFQ design and the summary of the initial tests with the prototype module are also  presented elsewhere \cite{RFQModule0}.

\section{Ion Source}
The ``Ion Source'' is a electromagnetic device is used to produce a particle beam of atomic nuclei. Ion Sources are the first component of any accelerator that uses different ions, from hydrogen to lead. 
Since their invention in 1946, the RF ion sources are preferred over those working via cathodic electron emission due to their longer lifetimes and lower maintenance requirements resulting from their thermionic electrode-less operation. Electrode-less operation becomes particularly important in the case of plasma of corrosive gases such as oxygen. Currently, RF ion sources are widely used to produce high-intensity ion beams in laboratories around the world, such as CERN\cite{CERN-IS}, IMP (China)\cite{IMP-IS}, 
PSI (Switzerland)\cite{PSI-IS}, SNS (US)\cite{SNS-IS} etc.  JPARC (Japan)\cite{JPARC-IS} is employing a solution featuring a LaB6 thermionic cathode in conjunction with an RF field to increase the beam current.  FNAL (US)\cite{FNAL-IS} and DESY (Germany)\cite{DESY-IS} ion sources work by bombarding the hydrogen gas using electrons originating from a magnetron.

In the RF ion sources, the electromagnetic field must be coupled to the plasma either inductively (ICP) using an antenna or capacitively (CCP) using parallel plates. The  CERN Linac4 H- ion source, Oak Ridge National Laboratory SNS ion source and PSI Hotlab are all using the ICP type ion sources \cite{CERN-IS, SNS-IS, PSI-IS}. 
Our group has previous experience with the ICP type ion source which usually requires water cooling for the power transmitting coil \cite{SPP-IS}.  

A third possibility for channeling the electromagnetic field towards the gas to be ionized is to use a waveguide to separate the emitting antenna (usually a microwave source) and the plasma. This is in fact the idea behind the so-called microwave ion source which originates from studies in Chalk River National Laboratory \cite{CRNL-IS},  Canada in early 1990s and also more recently aimed for the European Spallation Source\cite{ESS-IS}.  Such ion sources are characterized by a very long lifetime, of the order of a year or more. Another benefit is the smaller ionization chamber with respect to the previously discussed setups. The plasma is confined using magnets on both ends of the ionization chamber and the magnetic field properties define the operation mode. 
For example, if both the axial and radial magnetic field magnitude match the ECR conditions, the same particles can be ionized multiple times. Therefore these ion sources are typically used in heavy ion machines (e.g. CERN Pb beam). 
If a simpler magnetic field profile (axial only) and off the ECR resonance value is used, ionization usually occurs only once, making the setup more suitable for proton beams. This operation mode is called a microwave discharge  and the KAHVELab proton source presented in this paper is a microwave discharge ion source  (MDIS). For more detailed reviews of ion sources see \cite{ISrev1, ISrev2} and the references therein.

\begin{figure}[h]
\centering
\includegraphics[scale=0.30]{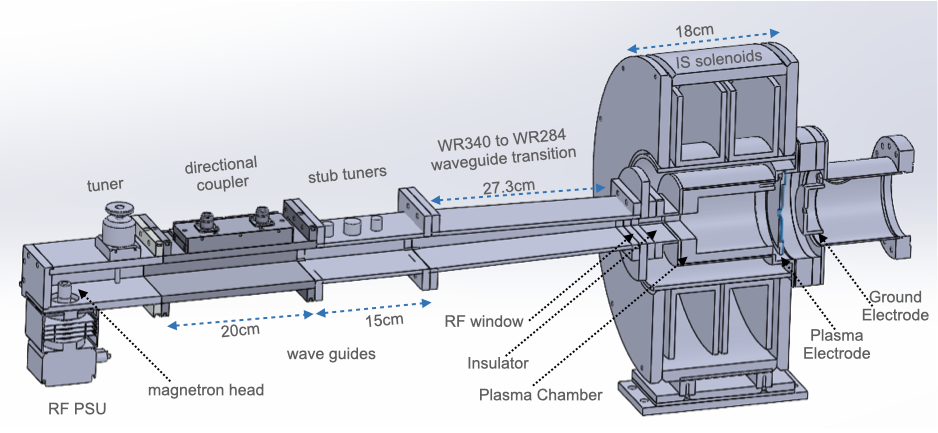}
\caption{The Microwave Discharge Ion Source design}
\label{fig:MDISdesign}
\end{figure}

The design view of the current MDIS at KAHVELab is an extension of the previous work presented in \cite{MISTestStand} focusing on making the system simple, reliable, low-cost and easy to use while increasing the extraction voltage from 10~kV to 20~kV. A step towards simplicity was to replace the previous 5 electrode extraction system (plasma electrode, ground electrode and Einzel lenses) with a new design, made with the IBSIMU program consisting of only 2 electrodes: plasma and ground. 
To increase the reliability of the MDIS, a better vacuum level was designed by correcting the connections around the plasma chamber mechanically and also by using thermally resistant O-rings. This improvement increased the vacuum level from 10$^{-5}$ mbar  to 10$^{-6}$ - 10$^{-7}$ mbar.  A number of control and monitoring upgrades, described below, were also designed for the RF section. The resulting MDIS design is shown in Fig.~\ref{fig:MDISdesign} as a cross sectional view of its three main components: The RF section, the plasma chamber and the extraction system.

The RF section starts with the leftmost device which is a commercial off the shelf RF power supply unit  that can be easily replaced. Currently, a 2.45~GHz magnetron (JENS JM002) with a 50~Hz repetition rate and about 800~W instantaneous power is used. 
The magnetron, its  high voltage transformer, and the wave guide access port are all hosted in a metallic cage  to prevent any RF wave leakage.  
As part of the upgrade, a tuner was added to the microwave head to achieve more precise tuning of the wave-guide RF transmission line. The RF transmission line consists of an RF source head converter, a stub-tuner, a WR340 to WR284 wave guide adapter, a high voltage breaker and a commercial vacuum tight RF window made from used quartz silica \cite{VacuumWindowScreen}. An RF directional coupler and its readout setup was added to the microwave transmission line between microwave head and stub tuner to determine the forward and reflected power values.
A wide band circulator with central frequency of 2.45 GHz is being designed and constructed for RF power supply protection and for removal for reflected power. It will be installed after successful testing. Currently, the heat due to impedance mismatch is removed via forced air cooling.  

The plasma chamber is of 90~mm diameter and of 100~mm length.  
The Hydrogen tank is stored under the MDIS setup and the gas flow is controlled by a remote controlled mass flow controller adjusted to a low value of 0.01 sccm (standard cubic centimeters per minute) for normal operation.
The plasma is generally confined using electromagnets which also ensure the ECR condition for efficient heating of electrons. However the high electric currents, and thus cooling water requirements for the solenoids render such a setup cumbersome. To overcome these problems a further upgrade of the ion source with permanent magnets was designed and later installed. The details of that study can be found in ref. \cite{IS-EPSProceeding}. 
However, this paper focuses on the commissioning work with the ion source using two solenoid electromagnets.
These solenoid magnets were capable of producing fields higher than ECR condition requirements and an iron cage was also added to render the magnetic field uniform. The details of this magnetic system design can be found in elsewhere \cite{MISTestStand}. Although no changes were made to the magnetic field design,  the electrodes, the insulator material and geometry have been upgraded for accommodating the increased extraction potential.

\begin{figure}[h]
\centering
\includegraphics[scale=0.45]{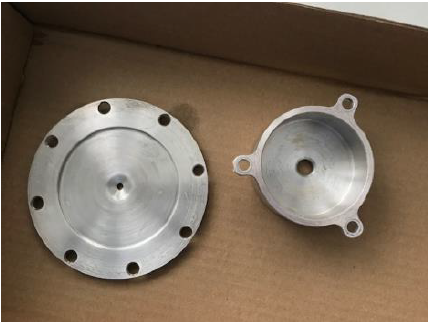}
\caption{The plasma (left) and ground (right) electrodes}
\label{fig:electrodes}
\end{figure}

\begin{figure}[h]
\centering
\includegraphics[scale=0.28]{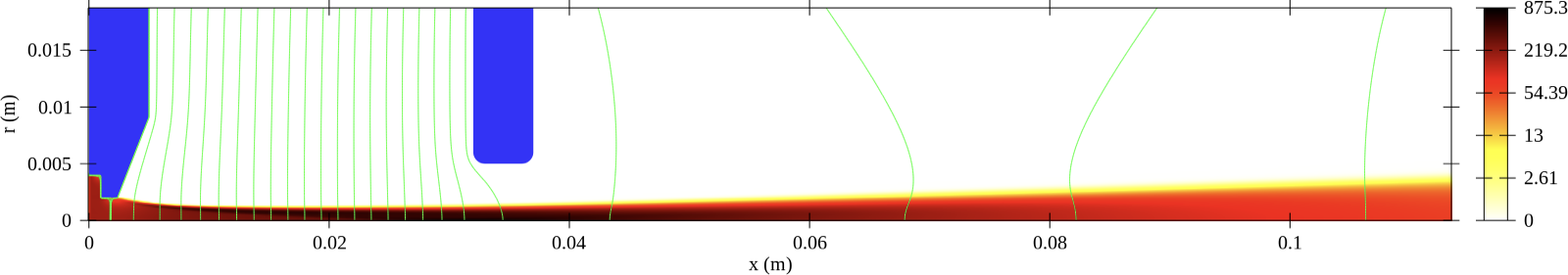}
\caption{Simulation of ions withdrawn from plasma using a two-electrode system (with IBSIMU)}
\label{fig:electrodesimulation}
\end{figure}

The extraction electrodes can be seen in Fig.~\ref{fig:electrodes} prior to installation. The plasma electrode has an aperture of 4 mm whereas the ground electrode’s aperture is 10~mm. 
The  Fig.~\ref{fig:electrodesimulation} contains a simulation of this setup, performed using the IBSIMU program \cite{ibsimu}. 
Matching between the plasma meniscus and electric potential can be optimized by adjusting the distance between the plasma and ground electrodes. 
The change in emittance for different plasma to ground electrode distances and for different $H^{+}$ ion current densities is also studied with the IBSIMU program. 
Simulation results show that for a proton beam current of 1.3~mA, the RMS emittance is 0.0254 $\pi$.mm.mrad .

The reliability issues that emerged during the commissioning were related to sparks between the high voltage of the plasma chamber and the ground connections at microwave input, Hydrogen gas pipe and beam extraction region. One particular difficulty was the unavailability of high purity alumina at an affordable price as high voltage insulation material since the thickness of the existing alumina insulator disks was determined as insufficient for the new 20~kV potential difference.

A low-cost alternative to alumina was investigated to solve the spark problem around the plasma chamber. A number of tests were conducted with acetal (commonly known as Delrin) and polytetrafluoroethylene (commonly known as Teflon) disks. During these tests it was found that Teflon easily accumulated dirt and dust, lowering its insulation capabilities. However this problem can be easily cured by simply washing the material with soap. Another disadvantage of these low-cost insulator candidates was that they can easily be scratched once again reducing their insulation abilities. Therefore it is recommended to use non-metallic wrenches and screwdrivers when working with these materials. After the evaluation, Teflon was selected due to its low cost, high breakdown voltage (100 times of Delrin) and ease of machining. 
Consequently, a Teflon ring of 20~mm thickness was inserted at the extraction side to isolate the MDIS high voltage section from ground and also a Teflon insulator jacket was inserted between plasma chamber and the MDIS solenoids to protect the rest of the system from high voltage. The gas pipe spark problem was solved after a few tests with different pipe materials. The material that was finally selected is Perfluoroalkoxy alkane which is also known as Teflon-PFA. The pipes made from Teflon-PFA have high working temperature up to about 260 degrees Celsius. 
The pipe used in the setup has an inner radius of 4~mm and an outer radius of 6~mm.

The extraction electrodes were made of aluminium which showed some shape change above the 5~mA beam current, especially at the thin sections of the electrodes. In the present setup, with low beam current, no such shape changes were observed. Nevertheless, future plans to change the electrode material with heat resistive molybdenum are in place.

\section{LEBT}
\begin{figure}[h]
\centering
\includegraphics[scale=0.50]{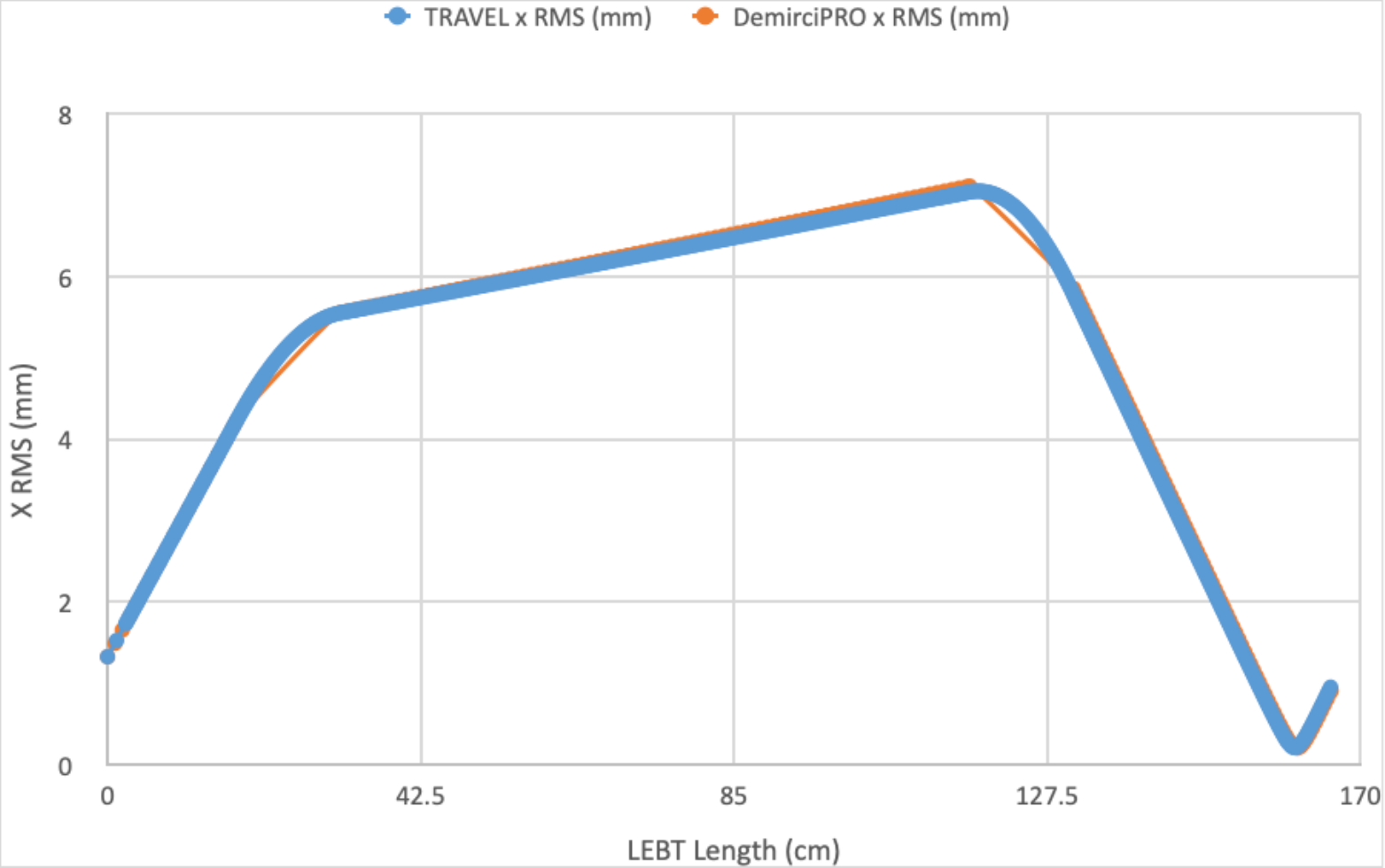}
\includegraphics[scale=0.50]{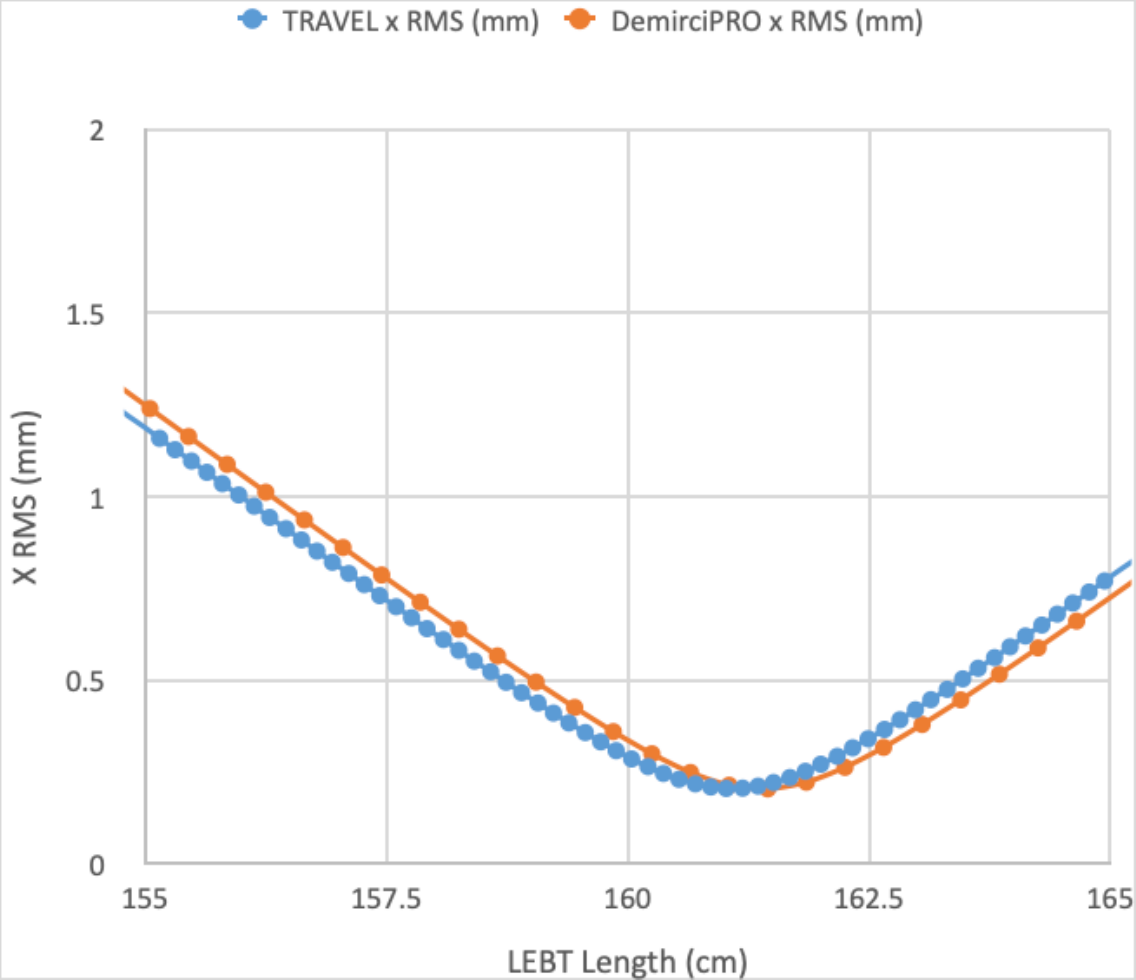}
\caption{Left: RMS beam envelope along the LEBT line for X axis, simulation from both TRAVEL and DEMIRCIPRO. Right: Zoom in view to the RMS minima}
\label{fig:LEBTLineDesign}
\end{figure}

The protons extracted from the ion source, are channeled to the accelerating cavity by using a Low Energy Beam Transport (LEBT) line.
The general requirements from the LEBT line is to deliver the ion beam into the next section, usually to the Radio Frequency Quadrupole (RFQ) matching maximally its acceptance requirements without loosing particles or leading to emittance growth. Typically this is achieved using either electrostatic or magnetic quadrupoles, or solenoids or Einzel lenses. 
Although the SNS LEBT line \cite{SNS-LEBT} uses two electrostatic Einzel lenses, ESS (Sweden)\cite{ESS-LEBT}, LINAC4 (CERN) \cite{CERN-LEBT}, TRASCO (Italy)\cite{Trasco-LEBT}, LEDA (US) \cite{LEDA-LEBT} and IPHI (France) \cite{IPHI-LEBT}, 
all proton machines, use a two solenoid setup for beam transportation together with the steerer magnets. 
Following the general trend in proton machines, the PTAK LEBT line also uses two-steerer (named St-1 and St-2) and two-solenoid (named Sol-1 and Sol-2) based setup for centering,  focusing and transporting the beam.
The TRAVEL\cite{travel} and DEMIRCIPRO \cite{demircipro} software programs were used to optimize the magnet positions and magnetic field strengths for acceptance match at the RFQ entrance. 
1869~G (Sol-1) and 1931.9~G (Sol-2) effective magnetic field strength values
were used as input values for both simulation programs. The behavior of the  RMS beam envelope  comparison between TRAVEL and DEMIRCIPRO are given in Fig. \ref{fig:LEBTLineDesign} left side, along the total LEBT length of about 165~cm. 
Both simulation programs are in agreement with each other and show no loss of beam throughout LEBT line. In the same figure right side, one can notice the RMS minimum occurs at about z=161 cm, where the RFQ will be positioned.

\begin{figure}[h] 
\centering
\includegraphics[scale=0.35]{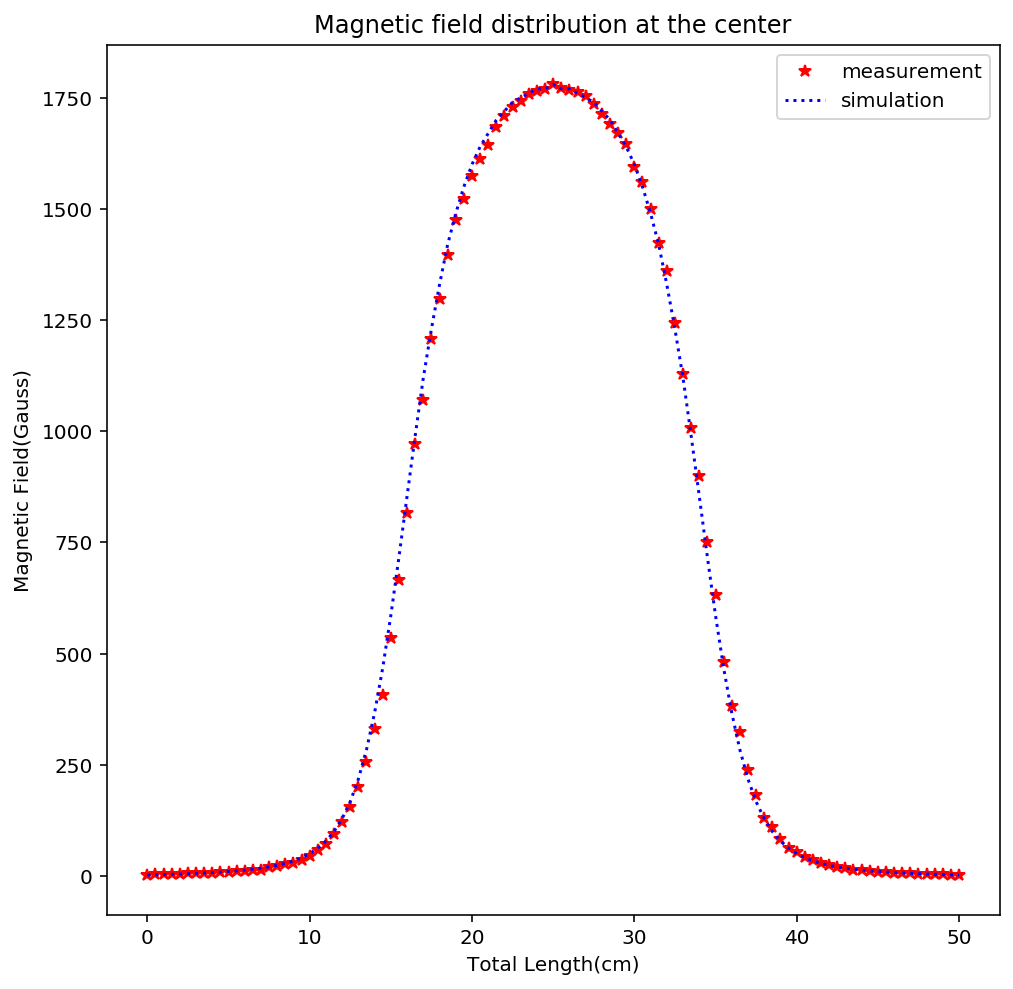} 
\includegraphics[scale=0.35]{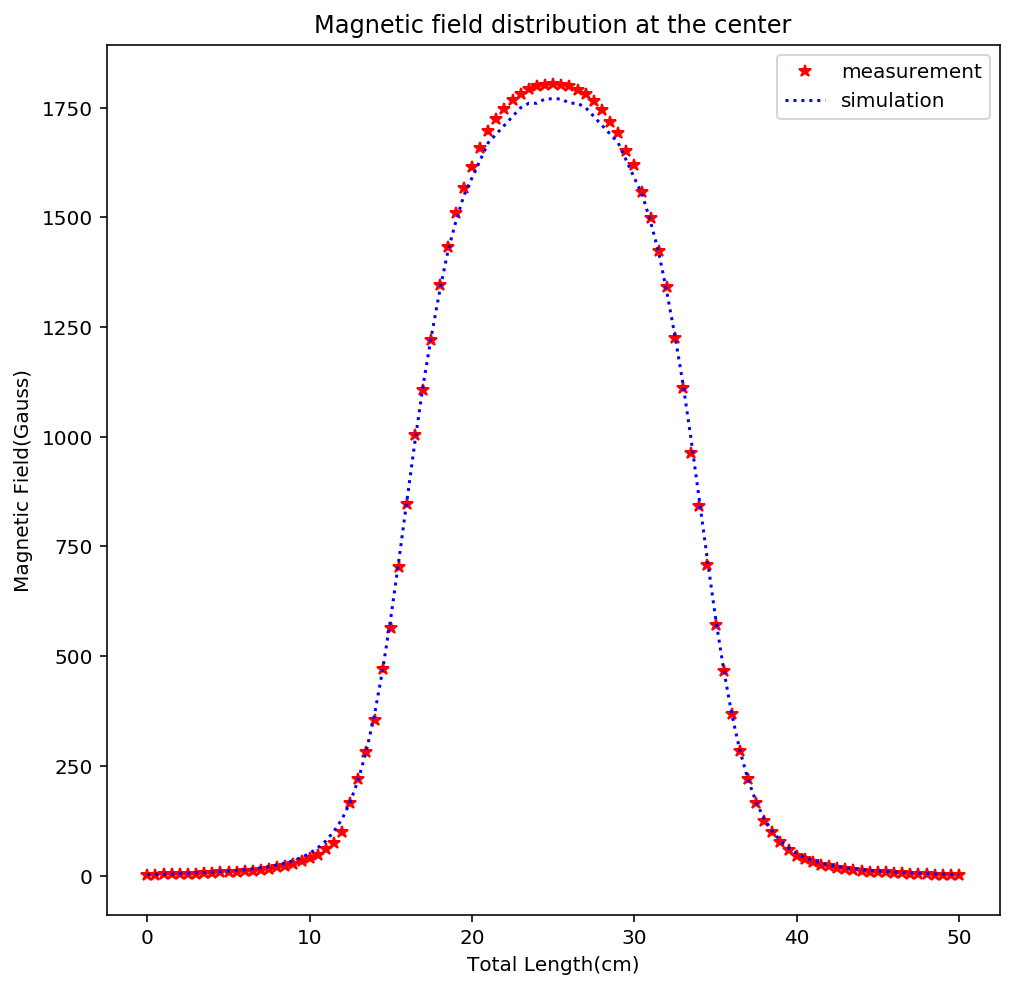}
\caption{Sol-1 (left)  and Sol-2 (right) magnetic fields at the center as compared to design values}
\label{fig:SolDesignMeasure}
\end{figure}

The solenoids of the LEBT line were designed as water-cooled electromagnets, each of about 22~cm length, with the center positions at about z~=~24.5~cm  at z~=~124~cm, respectively. The Poisson Superfish \cite{superfish} software program is used for the design and simulation. The flat top value of magnetic fields for both solenoids was set to about 1800~G for simulation and measurement comparison. The agreement between the designed and measured solenoid magnetic fields along the z-axis is shown in Fig.~\ref{fig:SolDesignMeasure} both solenoids. 

\begin{figure}[h] 
\centering
\includegraphics[scale=0.05]{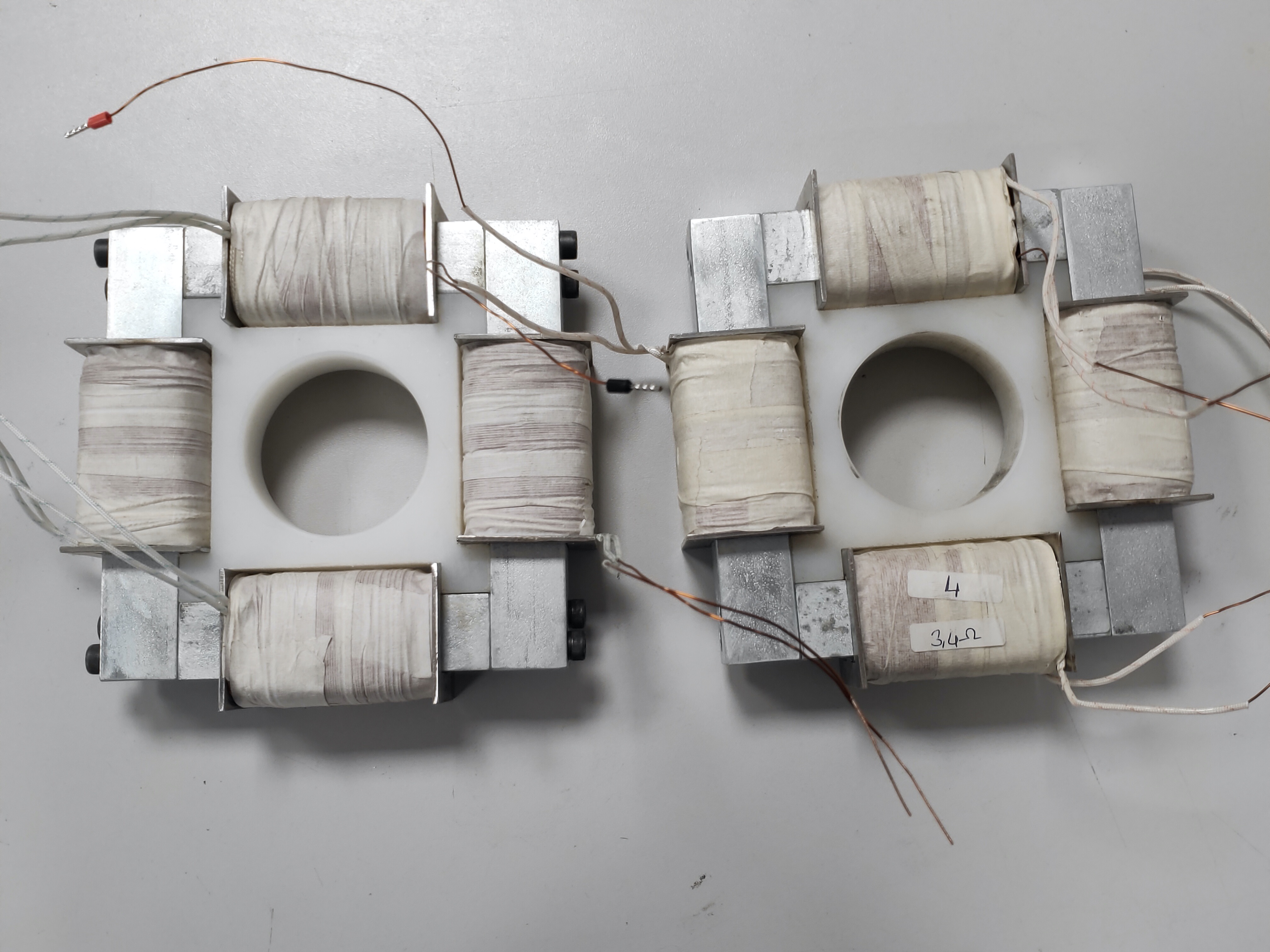}
\caption{Steerer Magnets for the LEBT Line.}
\label{fig:SteererMagnets}
\end{figure}

The steerer magnets are used to direct the beam within the beam pipe. 
Therefore two such magnets, one before the measurement station and one after were
designed to have iron cores and to produce about 63~G each. The magnet coils were designed with Poisson Superfish \cite{superfish} and the behaviour of the proton beam under these magnets was simulated with CST \cite{cst} programs. 
According to the design, the copper coil dimensions were determined as 64.60~mm in length, 29~mm in inner diameter, 39.60~mm in outer diameter, 75.60~mm distance between two rollers, 114.60~mm in length and 25~mm in width of stainless iron. 
The produced and assembled steerer magnets can be seen in Fig.~\ref{fig:SteererMagnets} .

\begin{figure}[h] 
\centering
\includegraphics[scale=0.3]{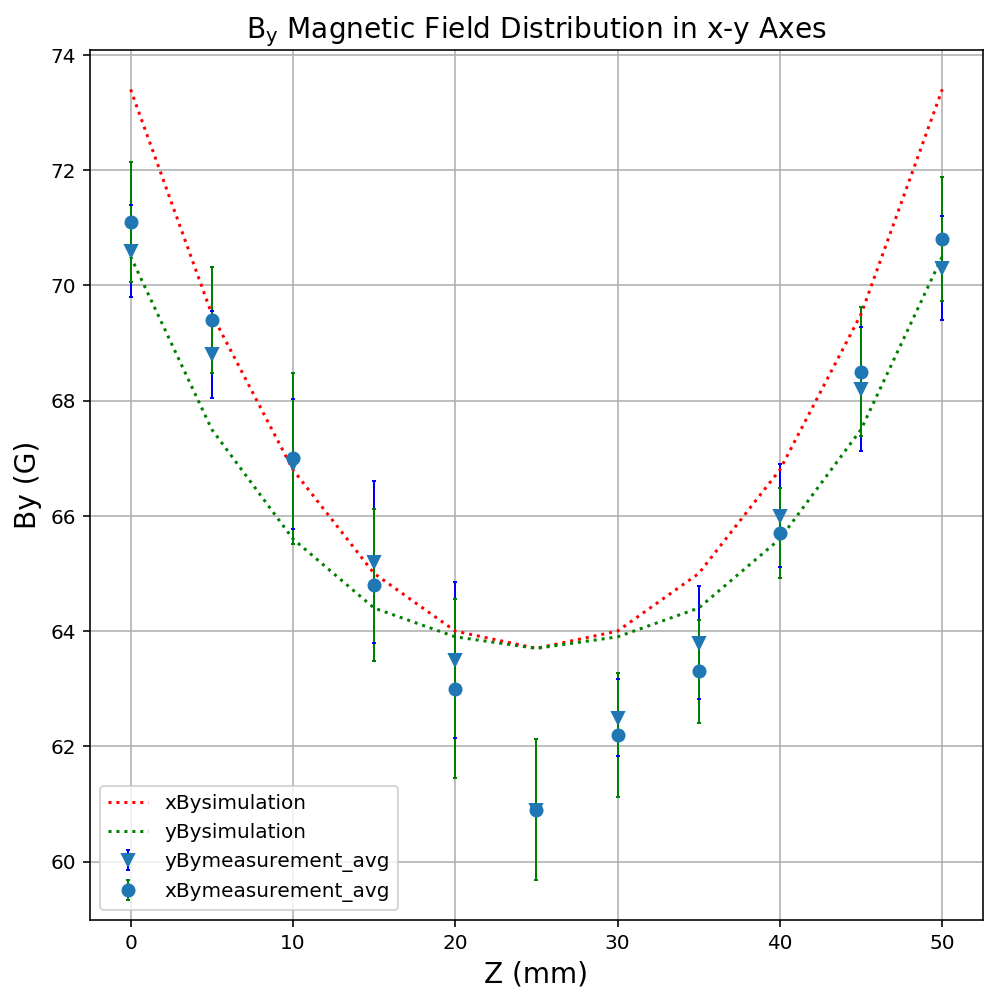}
\includegraphics[scale=0.3]{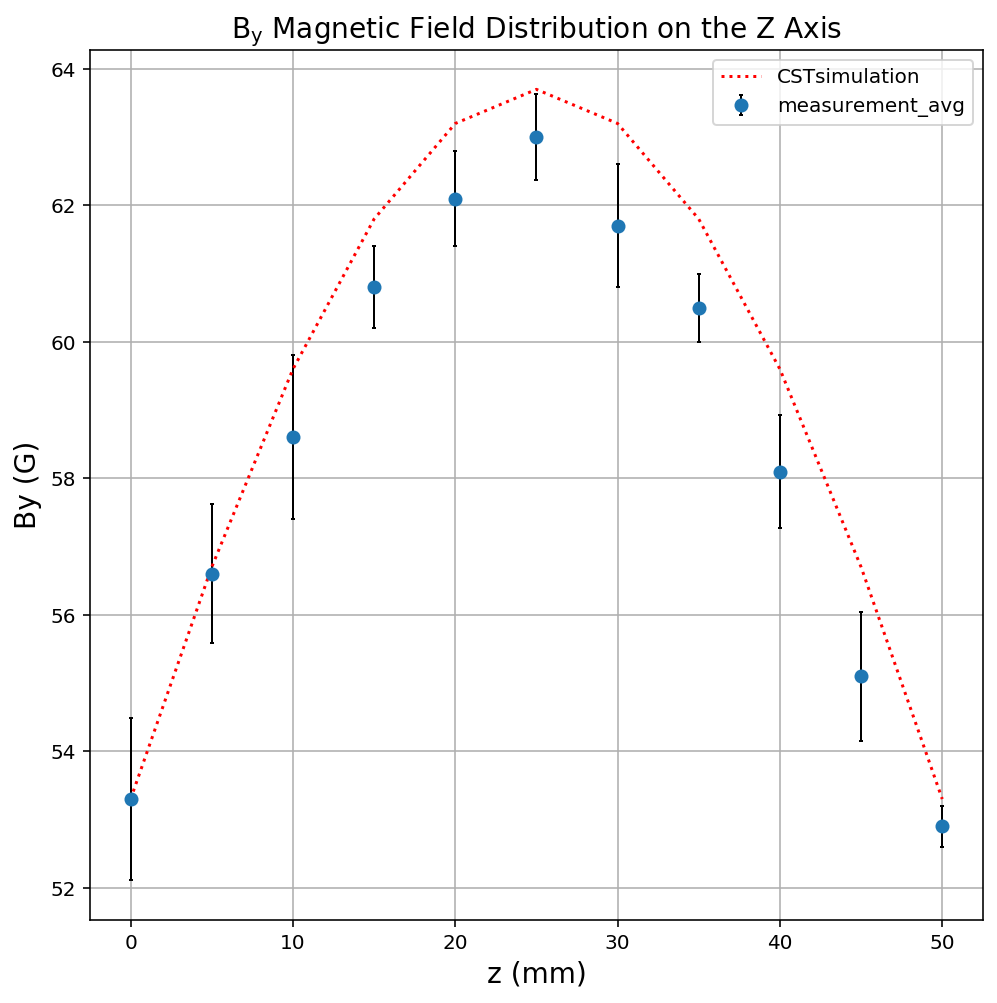}
\caption{Magnetic field measurements in the x,y (left) and z (right) by applying current to the coils facing each other (number 2 and 4).}
\label{fig:MagneticFieldMeasurements}
\end{figure}

The local producer was able to squeeze in about 584 turns of 0.76~mm thick wires in coil. Therefore with a 1A current the magnetic field peaks at approximately 63~G as expected. Magnetic field measurements were taken in the x, y and z axes by applying current to the coils facing each other, as shown in Fig.~\ref{fig:MagneticFieldMeasurements}.

\begin{figure}[h] 
\centering
\includegraphics[scale=0.11]{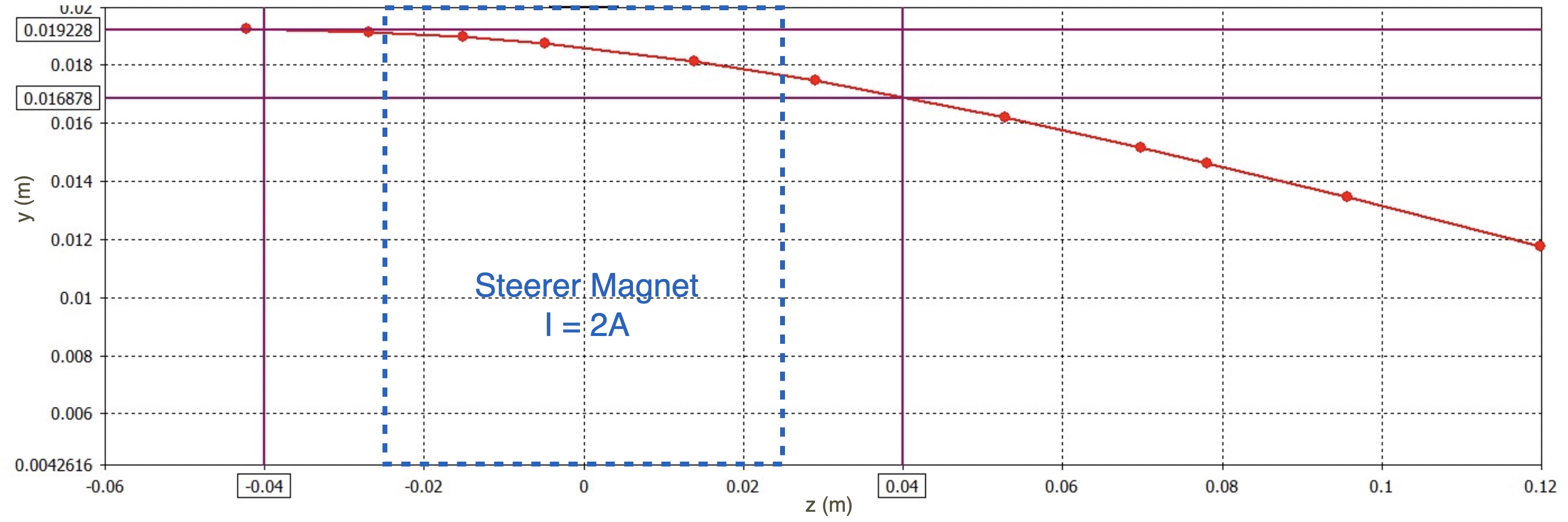} 
\caption{The effect of steerer magnets on the beam.}
\label{fig:BendingofSteerer}
\end{figure}

Using these steerer magnets, the 20 keV proton beam can be steered up to 2.35~mm horizontally or vertically at a distance of 40~mm from the magnet center as seen in Fig.~\ref{fig:BendingofSteerer}  where the blue rectangle in dashed line represents the steerer magnet, centered at z=0 in this plot. The deviation of the beam is also marked on the y-axis and can be calculated as $19.228-16.878 = 2.35$~mm.
The values and the plot are obtained from CST.

\subsection{Beam Diagnostics Station}
The previously discussed LEBT lines also contain beam diagnostics elements such as AC current transformers (ACCT), Faraday cups (FC), profile and emittance measurement tools. Sometimes the LEBT line might also contain one or more collimators to eliminate the beam halo or unwanted sections of the phase space in a controlled way. The design choices depends as usual on the past experience and available resources. For example the TRASCO's LEBT line contains a halo scraper collimator, a beam collimator and a FC between the solenoids. Their diagnostics are provided by the ACCTs, DCCTs and  CCD cameras. The ESS LEBT uses SEM grids after each solenoid for both beam profile and emittance measurements which also requires a slit type collimator. This is similar to CERN's Linac LEBT line which uses slit-FC method for emittance measurements placed inside a diagnostics box placed between the solenoids. IPHI does have a diagnostics station with an emittance meter and a FC after the second solenoid right before the RFQ.  The GSI (Germany) and BNL (US) emittance measurements are done via the pepper pot method which has the advantage of obtaining both horizontal and vertical components of the beam emittance simultaneously in a single measurement.

PTAK diagnostics design tries to pick up best properties of different existing designs while keeping the production cost low and leaving enough space for future devices. It therefore contains a diagnostics station between the solenoids which can measure the beam profile, emittance and current. The beam current measurement device is a Faraday Cup with an electron suppressor guard ring. The FC design was optimized with CST software \cite{cst} to minimize the number of secondary electrons escaping the cup. A phosphor screen is used to measure the transverse beam shape in both directions and also to calculate the transverse emittance by incorporating a pepper-pot plate. The details of the diagnostics station are discussed in \cite{MBOXPaper}. 

\begin{figure}[h]
\centering
\includegraphics[scale=0.50]{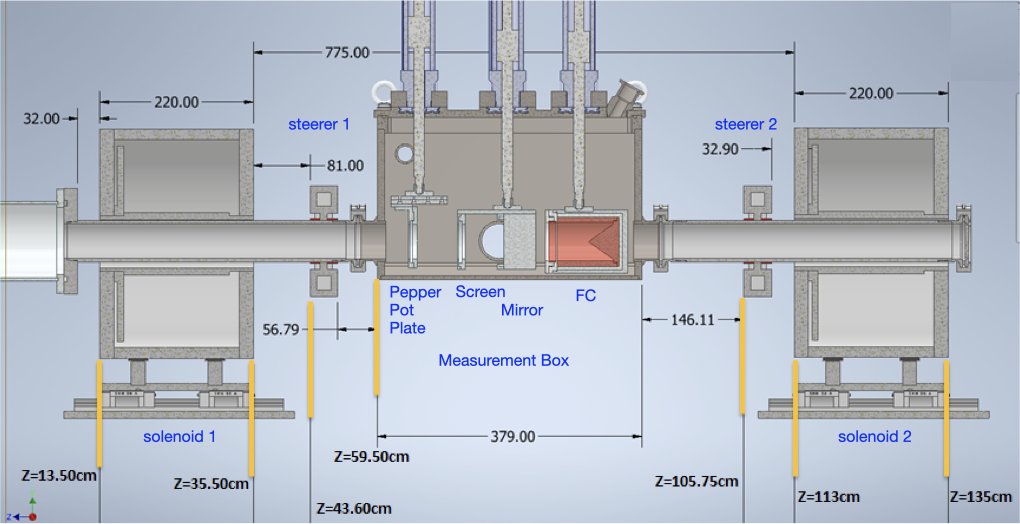}
\caption{Schematic drawing of LEBT line. Unspecified units should be assumed as mm.}
\label{fig:LEBTLineEng}
\end{figure}

The low energy beam diagnostics station is locally produced and installed between the solenoids, hence completing the LEBT line. Its engineering drawing is given in Fig.~\ref{fig:LEBTLineEng}. 
Enough space is left for planned future upgrades such as a AC current transformer that could be used for time of flight measurements.

\section{Status and Operations}
As of Q1 2022, the PTAK setup was installed excluding the RFQ to perform a number of commissioning runs. During these runs the proton beam was successfully transmitted to the end of the LEBT line. For the beam characterization, both the  Beam Diagnostics Station and an ad-hoc setup with a FC and a phosphorus screen were used.
There were a number of issues noticed during the commissioning studies.
For example in other beamlines, a high current power supply (100 A or higher) is used to power the focusing magnets, which results lower number of windings. Since it is difficult to properly prepare coils with large number windings this is a desirable situation. However the KAHVELab magnet power supplies have somewhat lower current outputs (up to 20 A’s) which required larger number of windings which was a challenge for the local industry.

The coil manufacturers in Istanbul, Turkey were able to wind 1682 turns on Sol-1 with a wire of 2.8~mm diameter and about 4000 turns on Sol-2 with a wire of 1.9~mm. It was possible therefore to power up the solenoids with the current sources at hand, however the temperature and current monitoring has become crucial to ensure the  stability of the magnetic field. Based on this example one must emphasize that in all experimental setups, even in those as modest as  PTAK, computer control, remote monitoring and operations are necessary for proper operation of the system; otherwise the repeatability of the measurements can not be guarantied.

Another issue that was seen during the commissioning was the warping of the 0.2~mm stainless steel pepper pot plate after multiple data taking sessions.  This problem was solved by sandwiching the plate between two  0.5~mm thick aluminum sheets to keep it straight and to ease the removal of  the deposited heat.

\subsection{Beam charge and duty factor}
\begin{figure}[h] 
\centering
\includegraphics[scale=0.50]{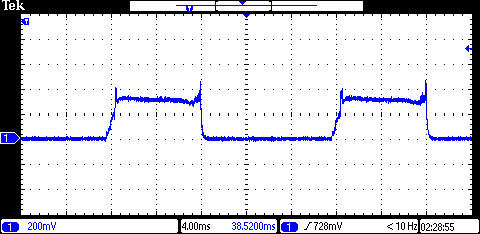}
\caption{FC signals: d.f. is found as 0.4 and the beam current as 0.03 mA}
\label{fig:FC}
\end{figure}

Two Faraday cups, one in the diagnostic station and another at the end of the LEBT line were used to measure the beam charge and the duty factor. A simple readout circuit with two  resistors of 10~k$\Omega$ each are used in series to connect the signal from FC to ground. The signal is read over one of the resistors to reduce the signal height, thus to protect the oscilloscope.
The result of beam current measurement using an oscilloscope that reads the FC located at z=87~cm  can be seen in Fig.~\ref{fig:FC}. It shows that the pulse width is 8~ms and the pulse period is 20~ms.
The duty factor can therefore easily be calculated as $d.f.= 8/20 = 0.4$. Similarly the instantaneous current is found to be 0.03~mA and the average current as 0.012~mA. 
This pilot beam with lower than the design current is obtained by reducing the magnetron input voltage using a manually controlled variac, by adjusting the very first tuner after the magnetron and also by reducing the gas flow rate to a minimum of 0.01~sccm. The rational behind the pilot beam is to test the whole system using a small charge and, once verified, to increase the beam current gradually. This is planned for after the completion of the RFQ installation. 

\subsection{Beam spot size}

\begin{figure}[h] 
\centering
\includegraphics[scale=0.08]{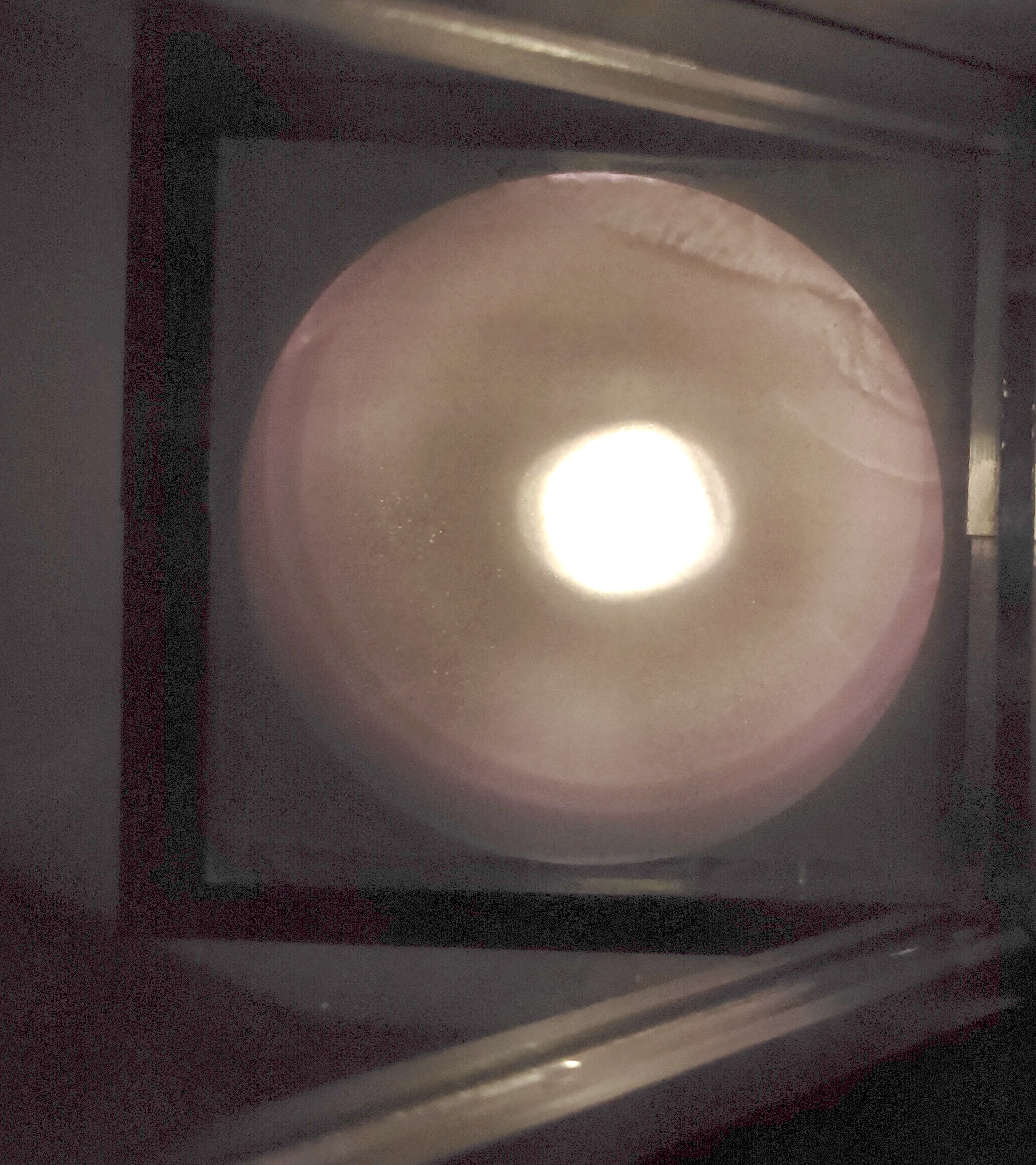}
\includegraphics[scale=0.205]{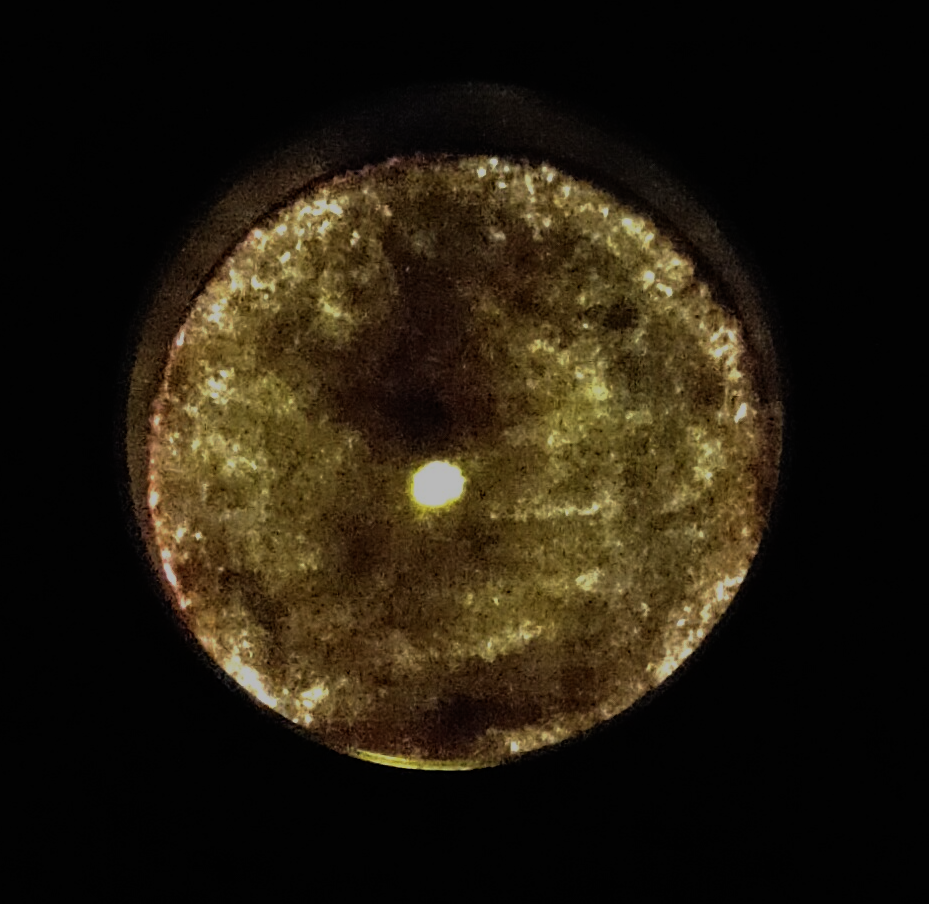}
\caption{Beam spot in diagnostics station (left) and after Sol-2 (right)}
\label{fig:BeamSpot}
\end{figure}

\begin{figure}[h] 
\centering
\includegraphics[scale=0.3]{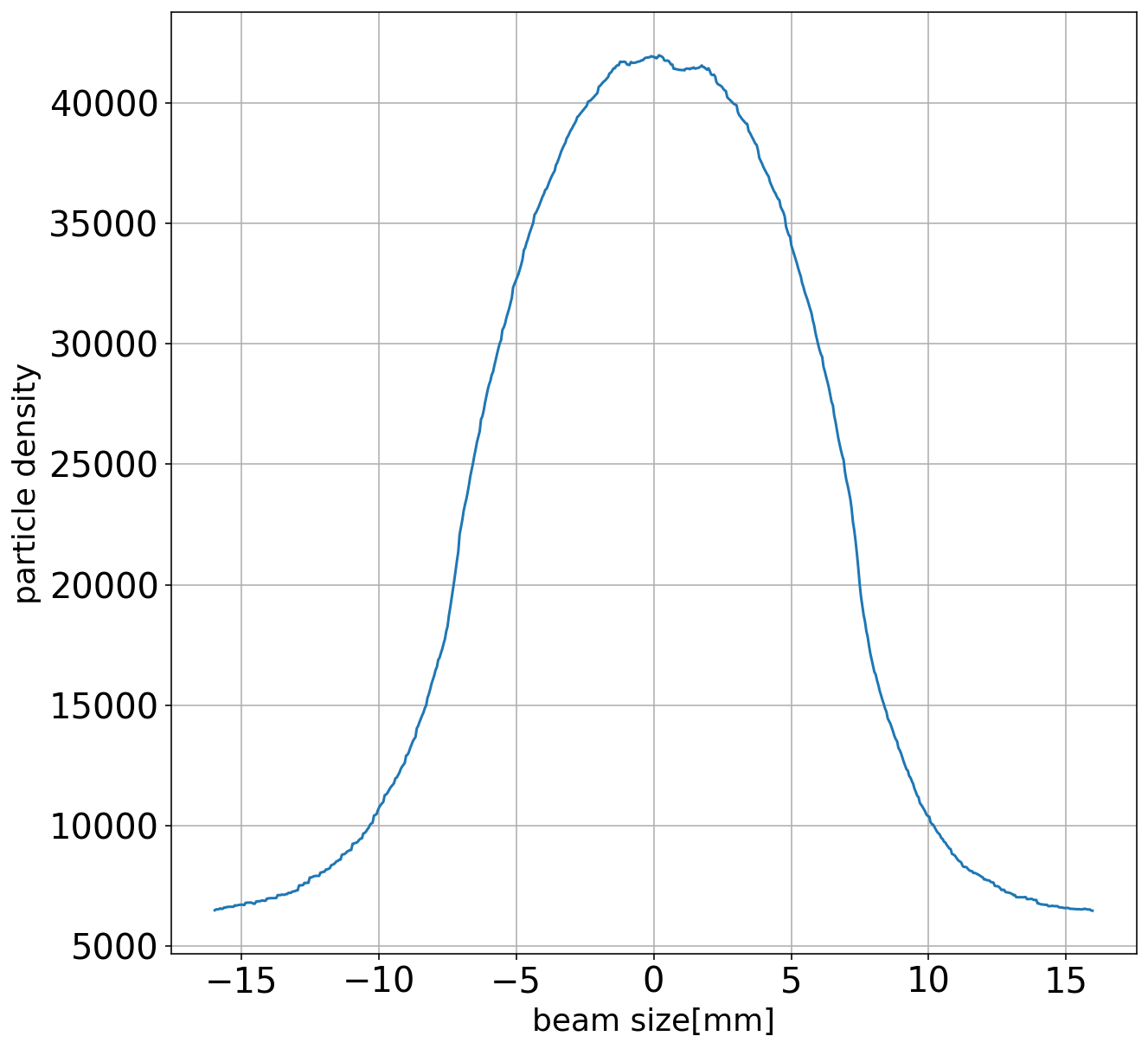}
\includegraphics[scale=0.3]{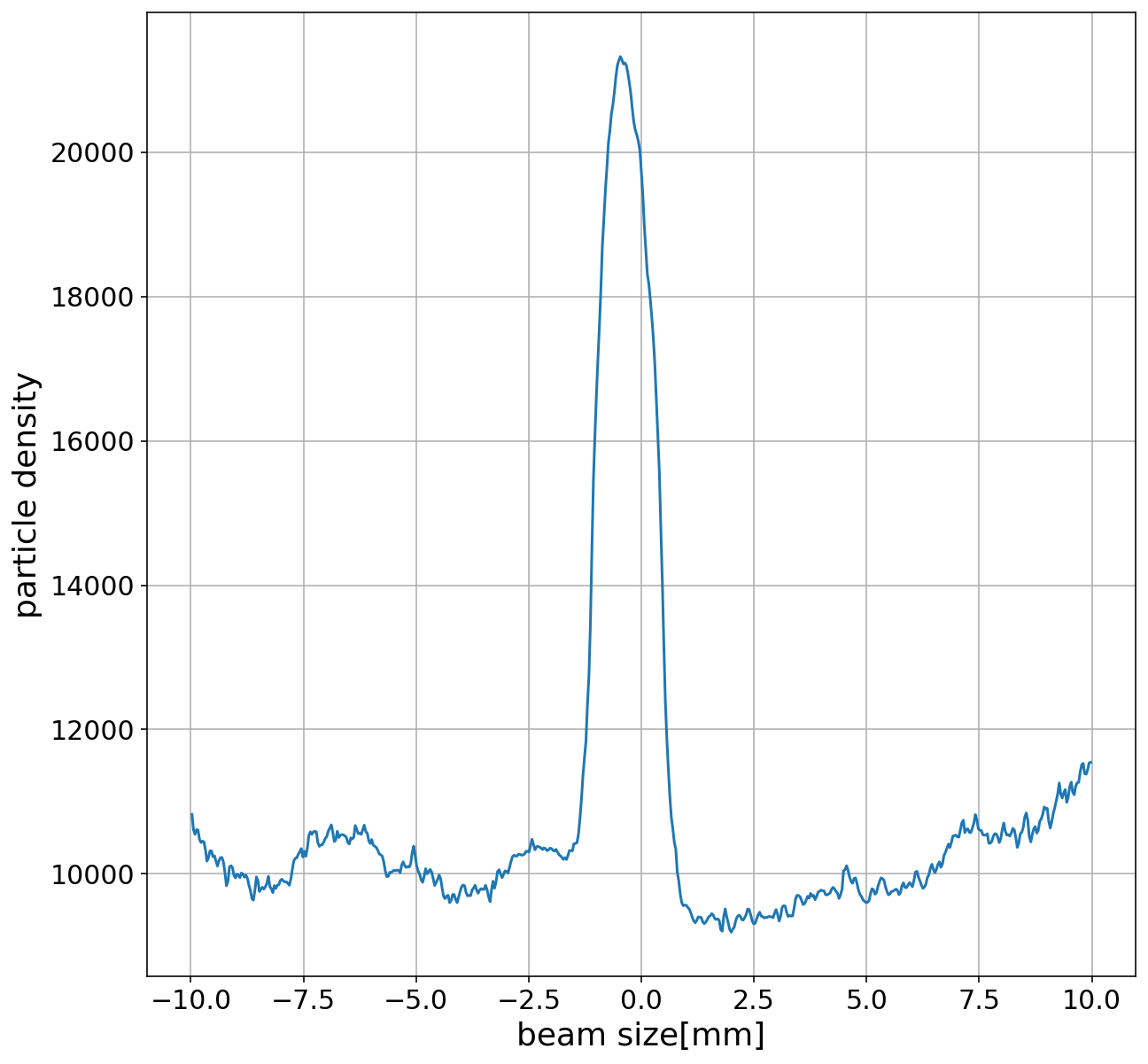}
\caption{The projection of beam size on the x-axis in diagnostics station (left) and after Sol-2 (right)}
\label{fig:BeamSize}
\end{figure}

The beam profile was measured in two locations: in the diagnostic station and after the second solenoid magnet (see Fig.~\ref{fig:LEBTLineEng} for detailed geometry). The beam image photos shown in Fig.~\ref{fig:BeamSpot} were taken using locally built phosphorus screens. A 300$\mu$m thick glass with $60\times60$ mm dimensions was coated with fluorescent powder to create this cost effective sensor. 
The photo on the right was obtained using an early prototype containing a gluing agent which, in time, deteriorated the image quality.
Inside the diagnostic station, a mirror mounted at a 45-degree angle behind the phosphor screen projects the image of the beams through the vacuum window into the camera. Therefore one can also see the mirror holding frame in the photo on the left side. 
The single image captured on these screens provide both X and Y profile information. The photos were digitized, converted into histograms and analyzed using a short computer program, written in Python.  Its results for x direction (as an example) can be found in Fig.~\ref{fig:BeamSize} for both images. For the image after sol-1, one can see that the beam is well centered and has a flat top central region. The beam diameter is calculated by finding the FWHM corresponding to about 15~mm. The beam size in y direction is slightly smaller corresponding to about 12mm.
The image obtained after sol-2 is more noisy and has a sharp peak at the center representing the focused beam aimed for the RFQ input. The FWHM method yields about 2~mm for both x and y beam diameters.

\subsection{Beam emittance}
\begin{figure}[h] 
\centering
\includegraphics[scale=0.20]{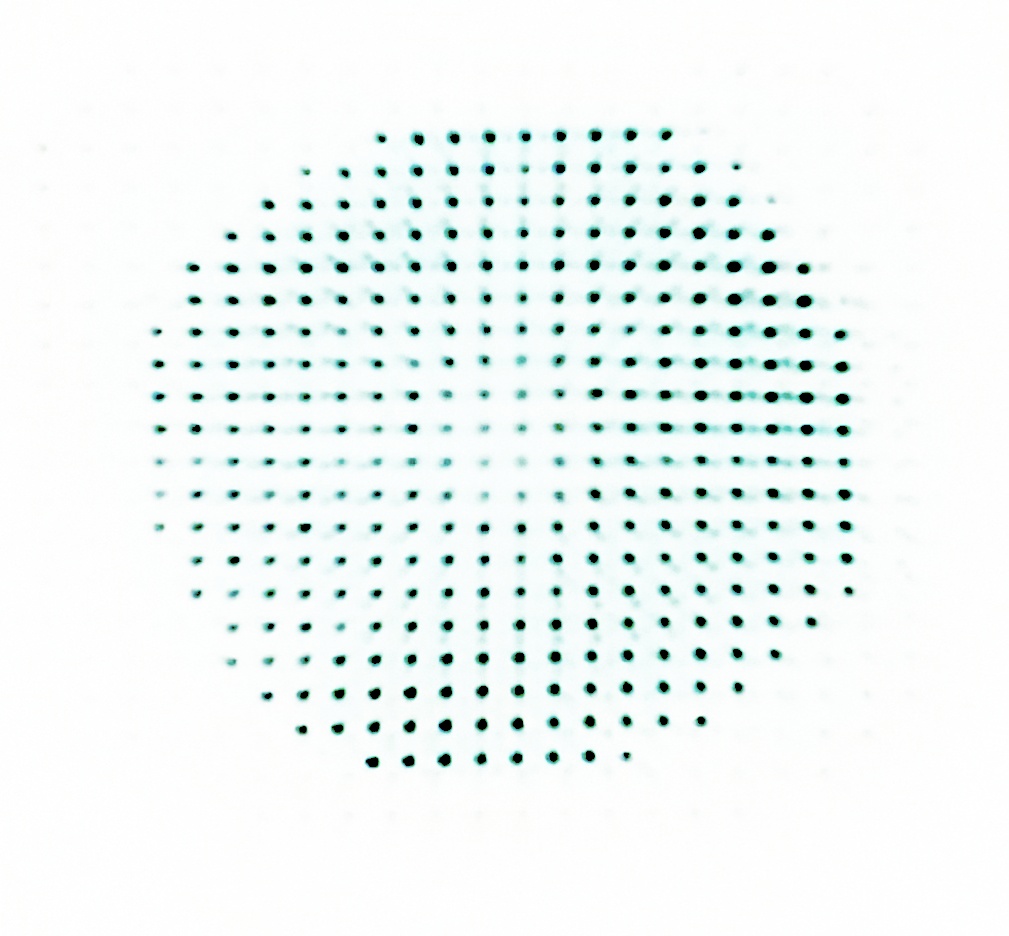}
\includegraphics[scale=0.20]{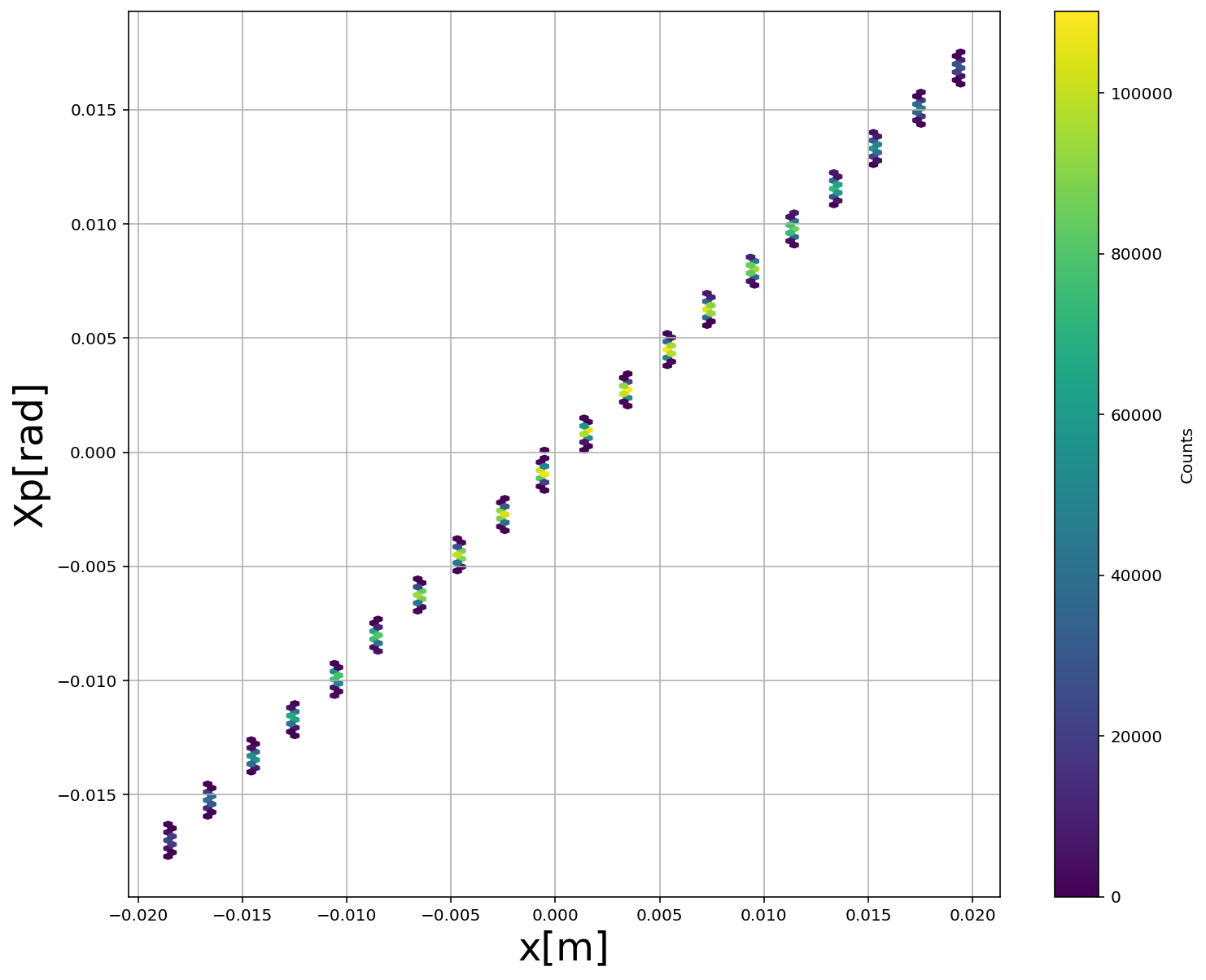}
\caption{Beam photo after pepper pot plate (left) and x emittance analysis result (right) }
\label{fig:BeamEmit}
\end{figure}

The beam emittance is measured inside the diagnostics station using the pepper-pot method \cite{PPMethod}. The pepper-pot plate is made of a $250~\mu m$ thick stainless steel and acts as a mask. Pinholes have a diameter of about 100 $\mu$m are spaced 2~mm horizontally and vertically and cover an area of $50 \times 50$~mm. The plate is sandwiched between two aluminum support frames of 500 $\mu$m thickness each. These frames are used for both thermal cooling and also to prevent any deformations that may occur on the surface of the perforated plate. A photo of the beam after the pepper-pot plate is presented in Fig.~\ref{fig:BeamEmit} left side. Such a photo can be analyzed to find the beam emittance and the Twiss parameters representing the beam.  Such an analysis software was developed locally in Python.
Although the details are given elsewhere, an example output for x-direction is shown in the same figure right side. The resulting RMS emittance is found as $\epsilon_n$=0.029 $\pi$~mm.mrad in agreement with the expected value of $\epsilon_n$=0.0254 $\pi$~mm.mrad obtained through simulations performed using IBSIMU\cite{ibsimu} and DEMIRCIPRO\cite{demircipro}. 
A summary of beamline simulation and measurement comparison results are given in Table \ref{tab:LEBTresults}.

\begin{table}
\caption{LEBT line simulation and measurement results. Emittance and Twiss parameter values are measured using the pepper pot plate at $z=62$~cm.}
\begin{center}
\begin{tabular}{|c| c | c|c|}
  \hline
  Parameters & Simulation & Measurement x (y) & Unit \\ \hline 
  beamsize in MBOX & 14 & 15 (12)                & mm  \\  \hline
  $\epsilon_{norm} $ & 0.031 & 0.029 (0.033)  &  $\pi$~mm.mrad \\ \hline
  $\alpha_{T} $ & -4.5 & -18.9 (-13.54) &  - \\ \hline
  $\beta_{T} $ & 1.33 & 2.13(1.83)      &  mm$ / \pi$~mrad \\ \hline
   beamsize after Sol-2  & 1.8  & 2 (2)   & mm  \\  
  \hline 
\end{tabular}
\label{tab:LEBTresults}
\end{center}
\end{table}

\subsection{Control and Monitoring}
\begin{figure}[h] 
\centering
\includegraphics[scale=0.25]{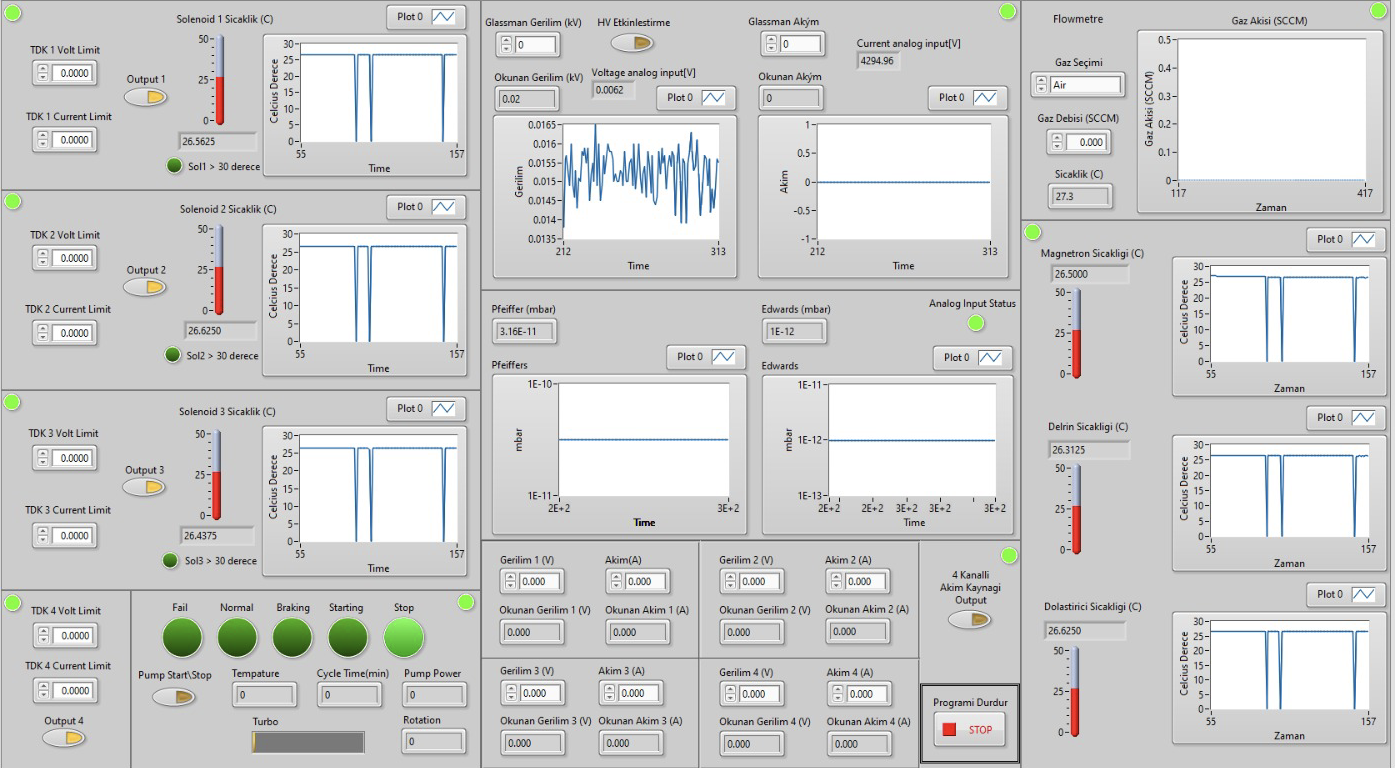}
\caption{PTAK setup control and monitoring GUI using LabVIEW}
\label{fig:gui}
\end{figure}

The control and monitoring (CM) system has to accommodate a high-voltage power supply, five current sources, two turbo-molecular vacuum pumps, two vacuum gauges, a flowmeter, three pneumatic cylinders and a number of temperature sensors. All devices controlled by a Graphical User Interface (GUI) written in LabVIEW's G language, shown in Fig. \ref{fig:gui}. The control and monitoring system, apart from the main PC, also includes one PLC (SiemensS7-1200 1215C) and two Arduino (Uno) micro controllers for improved equipment safety and also for limiting the load on the PC CPU.

One Arduino is used to read from six different sensors for measuring the temperature of solenoid magnets and other parts which are in direct contact the microwave power generator or close to the sections under high voltage. For example, the Ion source solenoid magnet has close contact with 20~kV high voltage.
Based on the readings a protection algorithm, implemented in the main CM software, can automatically shut down automatically the current sources which feed solenoid magnets. 
Therefore, a Bluetooth transmitter circuit has been added to the Arduino setup for wireless communication with the main PC. The wireless communication has the advantage of protecting the remainder of the control devices in case of a short circuit. 
The other Arduino is used to control the high voltage power supply which already has an analogue remote control port. It has two inputs, one for controlling the voltage and the other one the current, two analog outputs for monitoring those and finally a digital input to enable the high voltage.
LabVIEW GUI runs an algorithm in the background which controls current sources, turbo-molecular pumps, flowmeter, Arduinos and the PLC. Devices that have serial communication and are connected with RS232-USB converters to control computer. For other devices such as Agilent Twistorr Turbo-molecular pump which do not have an official LabVIEW driver, the necessary drivers were developed locally, therefore all such devices are governed by the CM program.
The PLC on the system is currently used to control the vacuum gauges and the pneumatic cylinders. The upgrade planned in 2022 is to move the control of both Arduino and LabVIEW systems from the LabVIEW based program to the PLC. In this scenario, LabVIEW will be used as a GUI to PLC operations.

\section{Outlook}
With these studies, the technical experience about ion sources, low energy beam transport line, solenoids, high voltage and vacuum has been accumulated. Maintaining the role of our university in the proton machine part of the Turkish Accelerator Center project at which we had an active role for many years, providing an opportunity for our students to work on accelerators needed in our country and obtaining the use of beam during two years for experiments and measurements are among the achieved aims of the study. While the analysis of the data collected during the commissioning runs continues, the old ion chamber was replaced with a new one where the magnetic field is provided with permanent magnets. The new system is to be commissioned in 2022. 
Another planned upgrade to be performed in this year is the installation of two ACCT stations for a time of flight measurement.

The PTAK beamline work is progressing smoothly, without any problems foreseen in the future. 
The local manufacturing abilities have proven to be adequate for all components produced so far. 
The production of the most critical component, the RFQ cavity is expected to finish in 2022. 
If there are no delays in production and commissioning, the first accelerated protons are expected by 2023.

\section*{Acknowledgements}
The main project is being supported by \.{I}stanbul University BAP grant 33250, as well as T\"{U}B\.{I}TAK grant 119M774. The PTAK line's RFQ and infrastructure has been supported by TÜBİTAK grant 118E838.

\end{document}